\pdfoutput=1
\documentclass[aps, reprint, superscriptaddress, nofootinbib, showpacs, floatfix]{revtex4-1}
\usepackage{amsmath, latexsym, amssymb, hyperref, graphicx, slashed, color, multirow}
\usepackage[T1]{fontenc}
\definecolor{nicered}{rgb}{.7,.1,.1}
\definecolor{nicegreen}{rgb}{.1,.5,.1}
\definecolor{darkblue}{rgb}{0,0,.5}
\hypersetup{colorlinks, citecolor=nicegreen,linkcolor=nicered, urlcolor=darkblue}

\newcommand\SEC[1]{\medskip\noindent{\sl\bfseries #1}}
\begin{document}
\addtolength{\belowdisplayskip}{-.5ex}
\addtolength{\abovedisplayskip}{-.5ex}

\title{Lepton Number Violation in Higgs Decay at LHC}

\author{Alessio Maiezza}
\email{amaiezza@ific.uv.es}
\affiliation{IFIC, Universitat de Val\`encia-CSIC, Apt. Correus 22085, E-46071 Val\`encia, Spain}

\author{Miha Nemev\v{s}ek}
\email{miha.nemevsek@ijs.si}
\affiliation{Jo\v{z}ef Stefan Institute, Ljubljana, Slovenia}

\author{Fabrizio Nesti}
\email{fabrizio.nesti@irb.hr}
\affiliation{Ru\dj er Bo\v{s}kovi\'c Institute, Bijeni\v{c}ka cesta 54, 10000, Zagreb, Croatia}

\date{\today}

\begin{abstract}
\noindent
We show that within the Left-Right symmetric model, lepton number violating decays of the Higgs
boson can be discovered at the LHC. The process is due to the mixing of the Higgs with the triplet
that breaks parity.  As a result, the Higgs can act as a gateway to the origin of heavy Majorana
neutrino mass.  To assess the LHC reach, a detailed collider study of the same-sign
di-leptons plus jets channel is provided. This process is complementary to the existing nuclear and
collider searches for lepton number violation and can probe the scale of parity restoration even
beyond other direct searches.
\end{abstract}


\maketitle

\noindent
The discovery of the Higgs boson~\cite{Higgs:1964ia,Weinberg:1967tq} allows to test the mechanism of
elementary particle mass generation at the LHC~\cite{Aad:2012tfa}. Compared to this success, the
problem of neutrino mass in the Standard Model (SM) appears acute. Neutrinos may be their own
antiparticles~\cite{Majorana:1937vz}, and lead to lepton number violation (LNV).  The canonical way
of searching for LNV, neutrino-less double beta decay ($0\nu2\beta$)~\cite{racahfurry}, can be
induced by light Majorana neutrinos or by new physics~\cite{feinbergbrown}.  The latter,
needed for neutrino mass, can be provided by the celebrated seesaw mechanism~\cite{Minkowski,
  MohSen, Yanagida, Glashow, GellMann}. In particular, Left-Right symmetric models
(LRSM)~\cite{Pati:1974yy}, designed to explain parity violation of weak
interactions~\cite{Senjanovic:1975rk}, embed naturally the seesaw~\cite{Minkowski, MohSen}. With the
left-right (LR) scale in the TeV range, $0\nu2\beta$ may be dominated by heavy Majorana neutrino
($N$) exchange~\cite{Mohapatra:1980yp, Tello:2010am}, which may become favored in view of the
cosmological bound on light neutrino masses.

A direct strategy for LNV searches at hadron colliders was suggested in~\cite{Keung:1983uu} by Keung
and Senjanovi\'c (KS)~\cite{GSreview}. The KS production of heavy Majorana neutrinos would reveal
LNV and relate directly to $0\nu2\beta$~\cite{Tello:2010am, Nemevsek:2011hz} and lepton flavor
violation (LFV)~\cite{Cirigliano:2004mv, AguilarSaavedra:2012fu}. The Dirac mass is
predicted~\cite{Nemevsek:2012iq} and may be tested at the LHC through LNV decays, uncovering the
underlying seesaw mechanism and connecting to electric dipole moments~\cite{Nieves:1986uk,
  Nemevsek:2012iq}. Indirect constraints~\cite{Beall:1981ze, Mohapatra:1983ae, Zhang:2007fn,
  Maiezza:2010ic, Bertolini:2012pu} played an important role and comprehensive
analyses~\cite{Bertolini:2014sua, Maiezza:2014ala} allow the LR scale well within the $\sim 6\,$TeV
reach of the LHC~\cite{Ferrari:2000sp}.

\begin{figure}
  \centering
  \includegraphics[width=.60\columnwidth]{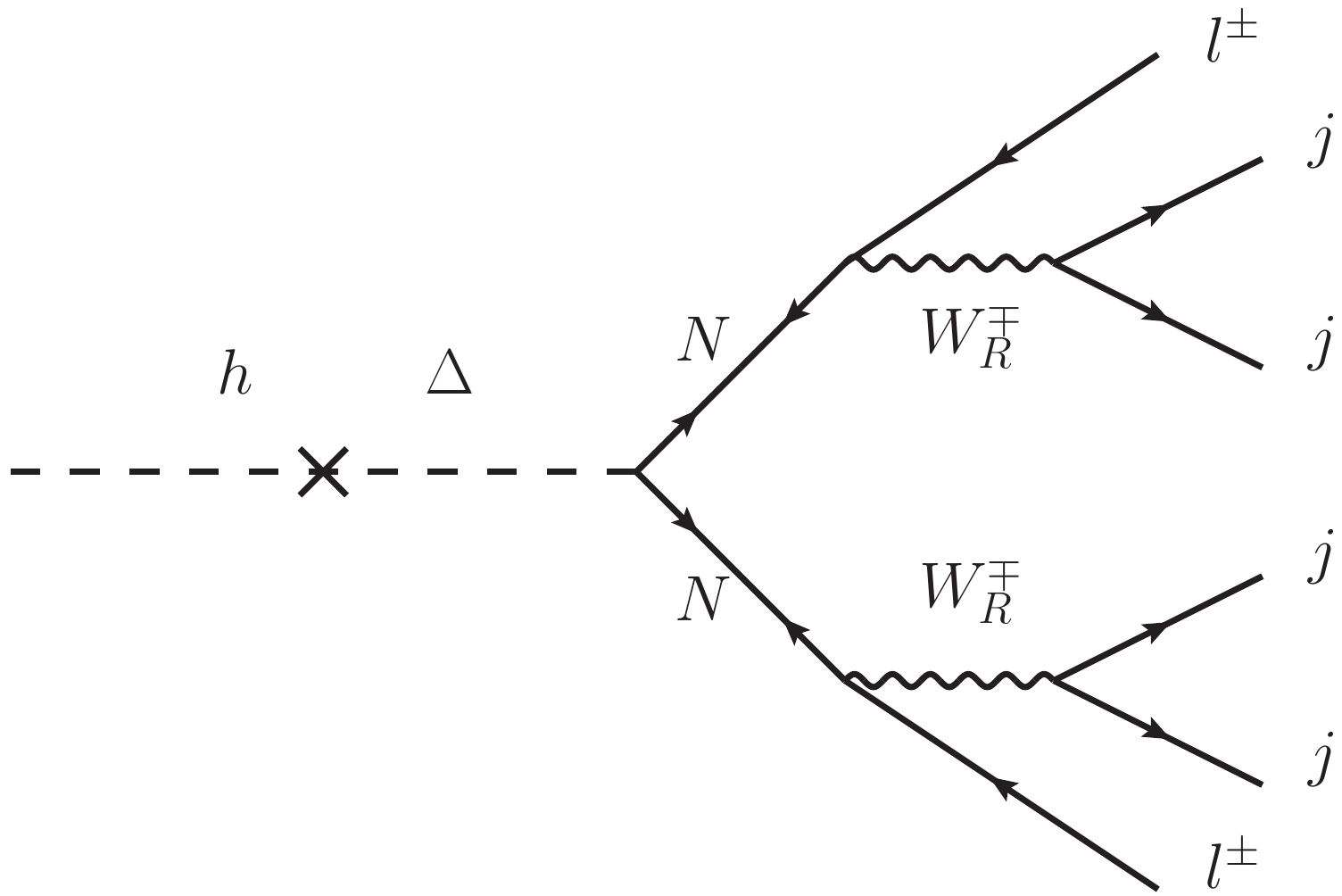} \vspace*{-.5em}
  \caption{Dominant diagram leading to LNV Higgs decay. \vspace*{-1.2em}}
\label{figLNVHiggs}
\end{figure}

In this Letter we show that LHC can probe a new channel, connecting Higgs physics to restoration of
parity.  We exploit the fact that the SM Higgs can have a sizeable mixing with the triplet that
breaks spontaneously LR symmetry and provides a mass to $N$. Through this mixing the Higgs can decay
to a pair of $N$~\cite{Gunion:1986im}. Therefore, it can probe their Yukawa couplings via a LNV
final state with two same or opposite sign charged leptons and four jets, as shown on
Fig.~\ref{figLNVHiggs}. The relevant range of $N$ masses typically leads to displaced
vertices. Higgs decay to RH neutrinos was mentioned in~\cite{Pilaftsis:1991ug} and studied
in~\cite{Graesser:2007yj} with effective operators, pointing out the LNV character and vertex
displacement.  Here, in the LRSM, the LNV decay is probing the origin of $N$ masses, just as the
standard decays test the origin of charged fermion masses. The Higgs can thus act as a portal to
LNV, complementary to $0\nu2\beta$ and the KS reaction.

To estimate the LHC sensitivity, we implement the model~\cite{LRMSMixModel}, perform a
simulation of the signal and the expected SM background, and devise cuts.
To further enhance the search, we simulate the displaced vertices from $N$ decay
and highlight their importance. Given the current limits on the Higgs
mixing~\cite{singletHiggs} (see outlook~\cite{Buttazzo:2015bka}), a discovery is
possible for a high LR scale, beyond the reach of other direct searches.

We conclude with a discussion on alternative models with potential LNV Higgs decays 
and a short outlook on the related search at $e^+ e^-$ colliders.

\SEC{Left-Right symmetry and Higgs mixing.}
Left-Right symmetric models~\cite{Pati:1974yy}, based on the gauge group $SU(2)_L \times SU(2)_R
\times U(1)_{B-L}$, contain a right-handed (RH) gauge boson $W_R$ and three RH Majorana
neutrinos. The scalar sector of the minimal LRSM~\cite{Minkowski, MohSen} features a complex bi-doublet
$\Phi \in (2_L, 2_R, 0)$ and a pair of triplets $\Delta_L \in \left( 3_L, 1_R , 2 \right)$, $\Delta_R \in
\left(1_L, 3_R, 2 \right)$:
\begin{equation}
\Phi = \begin{pmatrix}
  \phi_1^0 & \phi_2^+  
  \\
  \phi_1^- & \phi_2^0
\end{pmatrix}, \,\,
\Delta_{L,R} = \begin{pmatrix}
  \delta^{+}/ \sqrt{2} & \delta^{++}  
  \\
  \delta^0 & -\delta^{+}/\sqrt{2}
\end{pmatrix}_{L,R}.
\end{equation}

In the minimal LRSM, LR symmetry is restored at high energies. The potential exhibits spontaneous
breaking and since the original work~\cite{Senjanovic:1975rk} it has been the subject of several
studies~\cite{Basecq:1985sx, Gunion:1986im, originalHiggsmassLR, Zhang:2007fn,
  recentLRpotential}. We focus on the mixing between the triplet and the SM-like Higgs and display
the relevant terms:
\begin{equation} \label{eqScalarPot}
\begin{split}
  \mathcal{V} =& - \mu_1^2 (\Phi^\dag\Phi) - \mu_2^2 (\widetilde{\Phi} \Phi^\dag + {\widetilde{\Phi}}^\dag\Phi) - \mu_3^2 (\Delta_R^\dag \Delta_R^{})
  \\
  & + \lambda \, (\Phi^\dag\Phi)^2 + \rho \, (\Delta_R^\dag\Delta_R^{})^2 + \alpha (\Phi^\dag \Phi) (\Delta_R^\dag \Delta_R^{})\,.
\end{split}
\end{equation}
The trace on the parenthesis is implied and $\widetilde{\Phi} \equiv \sigma_2 \Phi^* \sigma_2$. The
results below hold for restoration of both generalized parity $\mathcal P$ and charge-conjugation
$\mathcal C$~\cite{Maiezza:2010ic}.%; a detailed discussion will be presented in~\cite{ourwip}.

The parameters $\mu$ are fixed 
%by spontaneous breaking 
in the usual way, $\mu_1^2=2 \lambda v^2+\alpha v_R^2$, $\mu_2^2=0$, $\mu_3^2=\alpha v^2+2 \rho
v_R^2$, and neutral scalars develop VEVs.  The LR-breaking scale is set by $\langle\delta_R^0
\rangle = v_R$ and electroweak symmetry breaking is completed by $\langle\phi_1^0 \rangle=v$. For
clarity we stick to the case where $\phi_2^0$ does not acquire VEV and we suppress higher $v/v_R$
terms.  In what follows the neutral scalars $\phi$ and $\delta$ are the fluctuations of
$\Re(\phi_1^0)$ and $\Re(\delta_R^0)$.

Expansion of the potential~\eqref{eqScalarPot} around the minimum gives the following mass matrix
for $\phi$ and $\delta$
\begin{equation} \label{eq:massmatrix}
  M^2 = 2 \begin{pmatrix}
    2 \lambda \, v^2 & \alpha \, v v_R 
    \\
   \alpha \, v v_R & 2 \rho \, v_R^2
   \end{pmatrix}.
\end{equation}
Its diagonalization leads to the masses of the physical particles $m_h^2\simeq4 \lambda
v^2-\alpha^2v^2/\rho$, $m_\Delta^2\simeq 4 \rho\, v_R^2$. Here $h = \phi \cos \theta - \delta \sin
\theta$ is identified with the SM Higgs boson and $\Delta = \delta \cos \theta + \phi \sin \theta$
with the further neutral state. Their mixing angle can be large,
$\theta \simeq  \left(\alpha/2\,\rho \right) \left(v/v_R \right)$.

Since $\delta$ is a SM singlet, mixing universally reduces the SM-like Higgs
couplings. Recent studies~\cite{singletHiggs} allow for $\sin \theta < 0.44$ at 2$\sigma$ CL, nearly
independently of the singlet mass.

\SEC{Heavy Neutrino from Higgs decay.}
%
%In the LRSM the above mixing leads to a new type of Higgs decay, with the fascinating possibility of
%probing LNV and neutrino mass generation via Higgs physics.
%
After spontaneous breaking, the Yukawa term $\mathcal L_\Delta = Y_N L_R^T \Delta_R L_R + \text{h.c.}$,
which couples the RH leptonic doublet $L_R$ to the triplet Higgs, generates the heavy neutrino mass
matrix. This is directly proportional to the LR scale
\begin{equation} \label{eqmN}
 M_N = 2 \, Y_N \, v_R\,,\qquad  M_{W_R} = g \, v_R\,,
\end{equation}
where $g=g_{L,R}$ is the $SU(2)_{L,R}$ gauge coupling constant.

To probe the spontaneous origin of $N$ mass, one should observe $\Delta \to NN$ decays and establish
that $\Gamma_{\Delta \to NN} \propto m_N^2$. While the production of $\Delta$ is small due to the
large LR scale, in the presence of $\theta$ the gluon fusion production appears.  Still, $\Delta$
may be heavy enough to be elusive, in fact from perturbativity of $\alpha,\lambda\lesssim 1$, one
finds $m_\Delta \lesssim 5\,\text{TeV} (0.1/\sin\theta)$.  More importantly, the SM Higgs can decay
to $NN$.  Thus the origin of neutrino masses may be probed by the SM Higgs boson.

\begin{figure}
  \centering
  \includegraphics[width=.8\columnwidth]{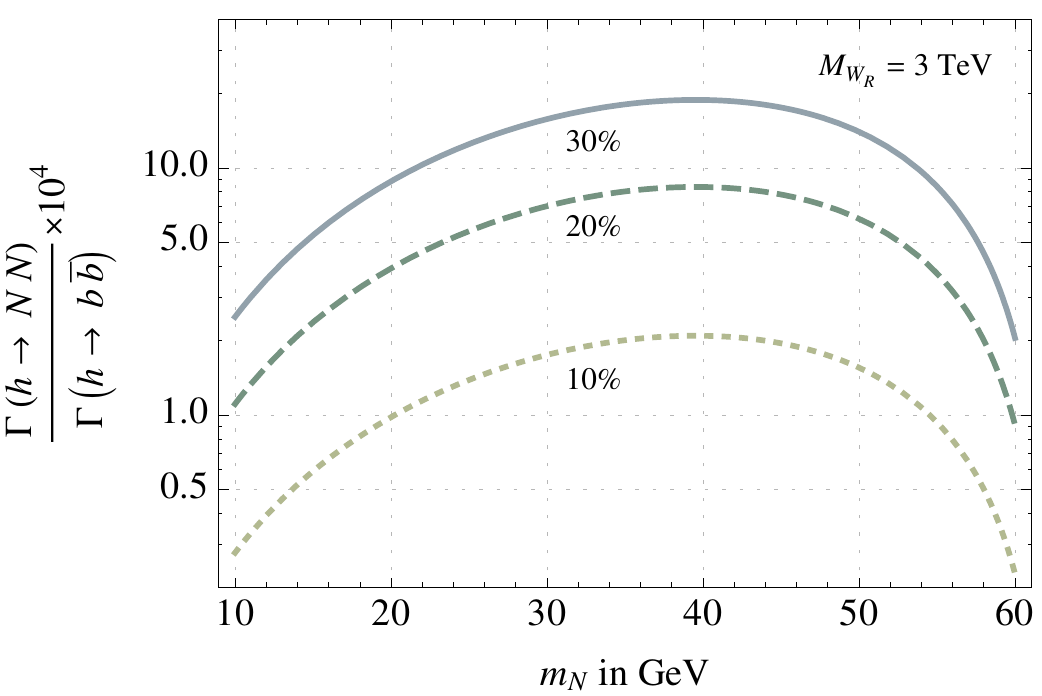}
  \caption{Decay rate of the SM Higgs $h\to NN$, normalized to the leading $h \to b \overline b$ channel.}
  \label{fig_ratio}
\end{figure}

It is useful to normalize this decay rate to the leading SM channel $h \rightarrow b \overline b$
% At tree level one gets
%
\begin{equation} \label{ratio}
  \frac{\Gamma_{NN}}{\Gamma_{b \overline b}} \simeq \frac{\tan \theta^2}{3} \left( \frac{m_N}{m_b} \right)^2 \left( \frac{M_W}{M_{W_R}} \right)^2
  \left(1-\frac{4 m_N^2}{m_h^2} \right)^{\frac{3}{2}},
\end{equation}
neglecting the $b\bar b$ phase space. Including the QCD running and NLO
corrections~\cite{Djouadi:2005gi}, the ratio %in Eq.~\eqref{ratio}
is enhanced by a factor $\approx 2$ and is shown on Fig.~\ref{fig_ratio}.

The number of $N$ pairs produced at the 13\,TeV LHC run with $100\,\text{fb}^{-1}$ luminosity is
simple to estimate. Taking the SM Higgs gluon fusion cross-section~\cite{Agashe:2014kda} $\sigma(gg \to
h) = 43.9\,\text{pb}$ and $\text{Br}(b \bar b) = 57\%$, one gets 500 (2000) $h \to NN$ events for $m_N
= 40\,\text{GeV}$ and $\theta = 10\% \, (20\%)$. This is sufficient motivation for an in-depth
collider study.

\SEC{Lepton number violating Higgs decay at the LHC.}
After pair-production from Higgs decay, each $N$ decays to a charged lepton and two jets
via $W_R$, with a RH charged current quark flavour structure essentially identical to the
left-handed one~\cite{Senjanovic:2014pva}. Due to the Majorana nature of $N$, $50\%$ of events will
result in a final state of two same-sign leptons and four jets with no missing energy, explicitly
signalling LNV.

To assess the LHC sensitivity, we extend~\cite{LRMSMixModel} the FeynRules~\cite{Alloul:2013bka}
implementation of the LRSM~\cite{Roitgrund:2014zka} to include the mixing together with Higgs gluon
fusion production.  Parton level events are simulated with MadGraph 5~\cite{Alwall:2014hca},
hadronized with Pythia 6~\cite{Sjostrand:2006za} and passed to Delphes 3~\cite{deFavereau:2013fsa}
for detector simulation. We use MadAnalysis~5~\cite{Conte:2012fm} for cuts and event
counting. Dedicated software extensions are implemented in each module to study the displaced
vertices.

%Below we discuss the physical characteristics of the signal and background sources, and then
%describe a cut strategy at detector level to optimize the final sensitivity.

The channel $h\to \ell^\pm \ell^\pm 4j$ carries plenty of physical information at parton level.  The
total invariant mass reconstructs the Higgs mass, while the $\ell j j$ invariant mass reconstructs
the $N$ peak. Tagging the lepton flavor identifies the RH analog of the PMNS mixing matrix and the
related Majorana mass matrix, as with the KS process~\cite{AguilarSaavedra:2012fu, vasquez}. Notice
that with such low $N$ masses, LFV constraints are easily satisfied and one may expect LFV in Higgs
decays.

Reconstruction at detector level is more delicate. The Higgs is produced with a boost $\gamma(h)
\sim 3$ at $\sqrt s = 13\,$TeV and the $N$ is further boosted if $m_N \ll m_h/2$. As a result the
two jets from $N$ tend to merge. In addition, the jets get closer to the charged lepton and typical
lepton isolation cuts may prevent its recognition. Furthermore, the distribution of transverse momentum
of the lepton (and the jets) peaks at a fairly low value of $m_h/6 \sim 20\,$GeV. Typical
detector simulation parameters forbid tight leptons with $p_T < 10\,$GeV, causing a loss of the
signal by a factor of 2. Still, the $N$ mass peak can be clearly observed in the $\mu j$ invariant
mass.

The fairly long $N$ proper decay length
\begin{equation} \label{eqNctau}
  (c \, \tau_N^0)^{-1} \simeq \frac{G_F^2 m_N^5}{16 \pi^3} \left(\frac{M_W}{M_{W_R}} \right)^4,
\end{equation}
characteristic for this portion of parameter space~\cite{Nemevsek:2011hz}, can lead to
displaced $N$ decay products. This ranges from sub-millimeter to a few meters,
depending on $m_N$ and $M_{W_R}$ and results in a striking LNV signature with two displaced vertices.

\SEC{Background estimation.}
Since lepton number is conserved in the SM, there is no background at parton level for this final
state. Nevertheless, there are three distinct ways in which background arises:

{\em 1.}\hspace{1ex} Electron charge mis-identification and secondary photo-production constitute a
background that is hard to understand in absence of real data. Since at this stage one cannot reliably
estimate this experimental effect, we study the muon channel free from such
issues~\cite{ATLAS:2014kca, Khachatryan:2015gha}.

{\em 2.}\hspace{1ex} The main prompt muon background comes from pair-production of electroweak gauge
bosons, in particular $WZ$, $ZZ$ and $W^\pm W^\pm j j$, and $t \bar t$ production. They
also contain non-prompt muons from mesons.

{\em 3.}\hspace{1ex} Significant background is due to non-prompt muons. This component, likely
dominant and not easy to estimate, is due to QCD jets when some hadron is mis-identified as a
muon. Although the mis-identification probability is small, the huge QCD production 
compensates~\cite{ATLAS:2014kca, Khachatryan:2015gha}. A realistic estimate will require a
knowledge of hadron mis-id within the real detector in the next LHC run. Nevertheless, previous
studies indicate this background behaves similarly to the $WZ+ZZ$ background (see supplement
of~\cite{Khachatryan:2015gha}). From that sample, we estimate the QCD mis-id contribution by multiplying
the $W Z + Z Z$ background by 2.5.

\SEC{Selection criteria and sensitivity.}
%
% We now turn to the event selection procedure.
We adopt the default Delphes 3 ATLAS card with muon
isolation parameters in agreement with~\cite{ATLAS:2014kca} and the anti-$k_T$ jet algorithm with
$\Delta R = 0.4$ and $p_{Tj min} = 20\,\text{GeV}$. We demand two same-sign isolated muons and no
other leptons, together with $n_j$ jets, where $n_j = 1,2,3$.  We require $\slashed{E}_T < 30\,
\text{GeV}$ and leading muon transverse momentum $p_T < 55 \text{ GeV}$. We demand the transverse
mass $m^T_{\mu \slashed p_T} < 30\,\text{GeV}$ and invariant masses $m_{\mu \mu} < 80\,\text{GeV},$
$m_{\mu \slashed p_T} < 60\,\text{GeV}$.  The impact of these selection cuts is shown in
Tab.~\ref{tabSigBack}. %on the event count

\begin{figure}[b]
   \vspace*{-1ex}
   \includegraphics[width=.96\columnwidth,height=.7\columnwidth]{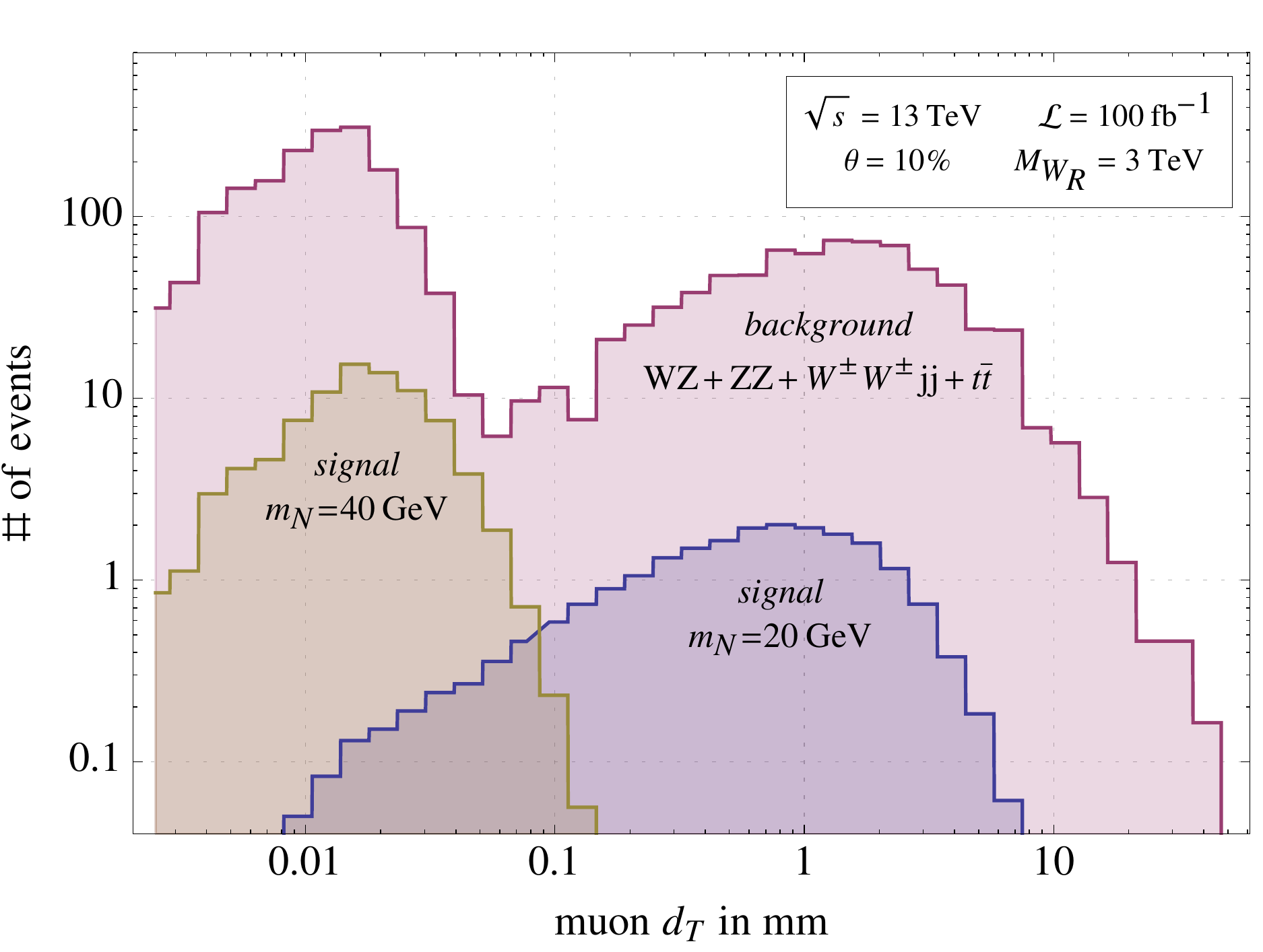} \vspace*{-1.5ex}
  \caption{Reconstructed transverse muon displacement after $\mu^\pm \!\mu^\pm \!\!+\! n_j$ event
    selection and before other cuts. \label{figDispMuonsT}}
\end{figure}

For both short and long lived $N$s the known decay length allows us to impose cuts on
the muon vertex transverse displacement $d_T$, shown on Fig.~\ref{figDispMuonsT}.  
%We simulate the displacement of signal and background, 
We smear the reconstructed vertex with the $p_T-\eta$ dependent resolution of 20--40\,$\mu$m, as
reported in~\cite{Aad:2011zb}. Since the typical background contains one prompt and one secondary
muon, it is effective to cut on both short and long $d_T$, on both muons. We employ a sliding window
cut, allowing events with $L/10 < d_T < 5 L$ and optimizing over $L$. This cut is expected to give
further control on the multijet QCD background.
% in addition to other known means, e.g.\ removing recognizable long lived
%mesons~\cite{Aad:2011zb}.

\begin{figure}[t]
  \vspace*{-1.2ex}%
  \includegraphics[width=\columnwidth]{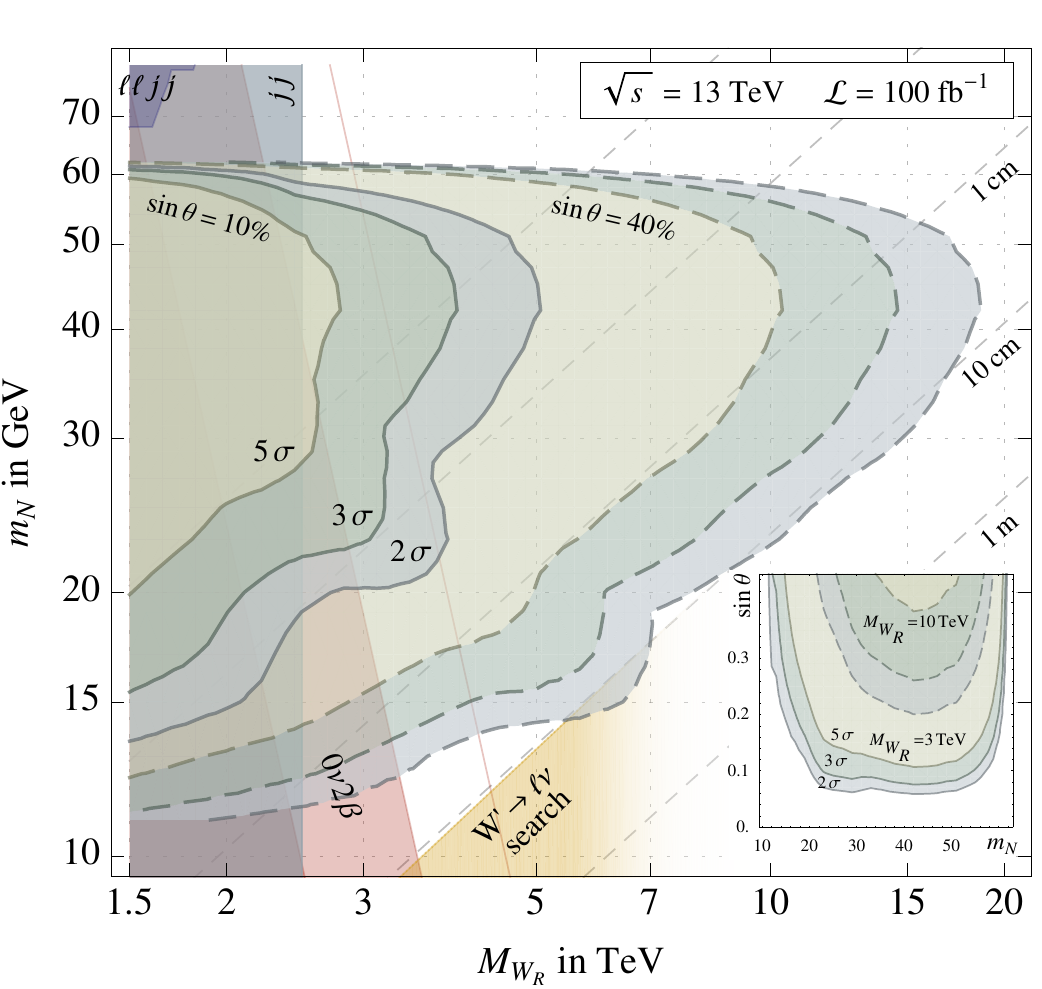}
  \caption{%
    Expected sensitivity in the $M_{W_R}-m_N$ plane, in the cases of small and large allowed mixing
    $\sin \theta$. %The inset shows the same in the $m_N- \sin \theta$ plane.
    The current bound from the KS channel~\cite{Khachatryan:2014dka} is shown in the upper left
    corner; the limit from di-jet search~\cite{Khachatryan:2015sja} as the vertical band. The region
    excluded by the GERDA I~\cite{Agostini:2013mzu} $0\nu2\beta$ search is the diagonal shaded
    region with a factor of 4 uncertainty from matrix elements~\cite{Barea:2012zz}, while the red
    line marks the sensitivity of GERDA II.  The grey dashed lines mark the lab frame decay length
    of $N$ from Higgs decay and the lower golden triangle the region where the length from $W_R \to
    \ell N$ decay exceeds 5\,m, enabling $W' \to \ell \slashed E$
    searches~\cite{Khachatryan:2014tva}.%
    \label{figWRsens}}
\end{figure}

The final sensitivity is shown in Fig.~\ref{figWRsens}, where a single $N$ with a $100\%$ Br to muons
is assumed. Given the allowed amount of Higgs
mixing $\sin \theta \lesssim 40\%$, this channel allows to test the LR scale beyond the
expected reach $M_{W_R}\sim6$\,TeV of direct searches~\cite{Ferrari:2000sp}.

\begin{table}[t]
  \begin{tabular}{| c  | c | c | c | c | c | c | c | }
  \hline \hline
  \multirow{2}{*}{Process} & \multirow{2}{*}{No cuts}  & \multicolumn{5}{c|}{Imposed cuts}
  \\ \cline{3-7}
   & & $\mu^\pm \!\mu^\pm \!+ n_j$ & $\slashed{E}_T$ & $p_T$ &	$m_T$ & $m_\text{inv}$
  \\ \hline
  $WZ$			& 2 M 	& 544 & ~143~ & ~78~   & ~40~ 	& ~20~
  \\
  $ZZ$			& 1 M	& 55	  & 29   & 16   & 12   	& 8
  \\
  $W^\pm W^\pm 2j$ & 389\rlap{$^\dag$} 	& 115 & 16   & 5     & 3  	& 1
  \\
  $t \overline t$ 		& 10 M\rlap{$^\dag$} 	& 509 & 97  & 40    & 22  	& 14
  \\ \hline
%  ~Signal (20)~		& 254 	& 11  & 11    & 10   & 9	& 8
%  \\
  ~Signal (40)~		& 543 	& 44 & 43    & 41   &  38 	& 37
  \\ \hline \hline
  \end{tabular}
  \caption{Number of expected events at the 13\,TeV LHC run with $\mathcal L = 100$\,fb$^{-1}$ 
   after cuts described in the text. The signal is generated with %$m_N = 20$ and
   $40\,\text{GeV}, \sin \theta = 10 \, \%$, $M_{W_R} = 3\,\text{TeV}$ and $n_j = 1,2,3$. 
   $^\dag$Here we restrict to $t \to b W$  and $W \to \mu \nu$.
   \vspace*{-1ex}
  \label{tabSigBack}}%
\end{table}

\SEC{Other models with LNV Higgs decays.}
The heavy fermions $S$ in standard type-I~\cite{Minkowski, MohSen, Yanagida, Glashow, GellMann} or
type-III~\cite{typeIII} seesaw couple to the Higgs boson via their Dirac masses $m_D$. If these were
rather large (unlikely in LRSM~\cite{Nemevsek:2012iq}) $h \to \nu S$ decay might be
observed~\cite{BhupalDev:2012zg} and via further mixing $m_D/m_S$, the LNV channel $h \to SS$ could
open up~\cite{Pilaftsis:1991ug}.  Below the $Z$ mass, where such two body Higgs decay is available,
constraint on the mixing~\cite{Abreu:1996pa, Khachatryan:2015gha} suppresses it beyond observation.

Further extension with a singlet~\cite{Shoemaker:2010fg} or spontaneously broken $B-L$ models can
generate an observable signal. %Our analysis applies to those cases as well.

Models with supersymmetric R-parity violation, alternative to the seesaw
mechanism~\cite{Grossman:2003gq}, lead generically to LNV.
%
% (both at colliders and at low
%energy via $0\nu2\beta$~\cite{Allanach:2009iv}).  
Same-sign di-lepton decay of the Higgs boson could follow from a fairly large $h \to \tilde \chi^0
\tilde \chi^0$ coupling, where $\tilde \chi^0$ decays to $\ell jj$ thanks to a non-zero $\tilde l q
q'$ term.  This scenario may deserve an updated study in light of recent limits.  In case of
slepton-sneutrino mass degeneracy, the LNV mode appears to be suppressed~\cite{Banks:2008xt}.

%Finally, in the context of a sequential fourth generation, decays to a pair of fourth generation
%Majorana neutrinos $h \to \nu_4 \nu_4$ was studied by~\cite{Datta:1991mf}. After the first run
%of the LHC, this framework seems to be disfavored.

\SEC{Outlook.}
In the SM, the Higgs mechanism provides masses to charged fermions and leads to a distinct
prediction of branching ratios. A completely analogous mechanism operates in the minimal
LRSM for $N$ and $W_R$~\cite{Minkowski, MohSen} where their masses determine the branching ratios of the
'right-handed' Higgs.  In this Letter we point out that the mixing between these two bosons leads to
$h\to NN$ and to the interesting possibility of LNV in Higgs decay.

The main results are summarized on Fig.~\ref{figWRsens}. For a large range of allowed $m_N$ and
$\theta$ values, the $h\to \ell^\pm\ell^\pm + jets$ channel allows to identify the RH neutrino mass
peak and can probe the LR scale even beyond the reach of $0\nu2\beta$ or collider searches.

Conceptually, this channel together with evidence of $\theta \neq 0$ can suffice to test the origin
of $N$ mass within this scenario. To check Eq.~\eqref{eqmN} one needs three independent
measurements. The event rate and invariant mass peaks give information on $\theta \times Y_N$ and
$m_N$, respectively. The decay length then constrains $M_{W_R}$ via Eq.~\eqref{eqNctau} and finally
a global fit on the Higgs data can provide $\sin \theta$.  If more than one $N$ were observed, this
task simplifies. To complete the understanding of neutrino mass origin, one would clearly like to
observe $\Delta$ and the associated gauge boson that provides the gauge symmetry protection.  
This channel would nevertheless allow to probe the polarization of $N$ decay, similarly to the KS
channel~\cite{Han:2012vk}, even if $W_R$ were out of reach.  A comprehensive study will be presented
in~\cite{ourwip}.

%From the collider perspective, the present analysis leaves room for improvement once the detector
%knowledge will be available in the next LHC run.  
A number of potential improvements can be
identified: i) %it seems feasible to include reconstructed 
adding muons with $p_T<10\,$GeV would double the signal;
%leading to a factor of two more signal events; 
ii) for short lived $N$, tight cuts on displacement could reduce further the QCD multijet background; 
iii) for boosted long lived $N$s, the muon tends to merge with the jet and one gets displaced jets.
Although a challenge, it may be possible~\cite{Schwaller:2015gea} to identify their
displacement.

We conclude by pointing out that $e^+ e^-$ colliders provide a particularly clean environment for
heavy neutrino searches~\cite{Blondel:2014bra}. For the LNV Higgs decay, the relative decrease in
production cross-section to $\sigma = 0.24\,$fb at $\sqrt s \sim 240\,\text{GeV}$ may be compensated
by lack of background (only $ZZ$ remains) and large luminosity
$\sim1-10\,$ab$^{-1}$~\cite{Gomez-Ceballos:2013zzn}. Conversely, a positive signal at LHC without the
associated $W_R$ discovery would make a case at a high energy hadron collider.

\begin{acknowledgments}

\noindent We thank G. Senjanovi\'c for useful discussions on the theoretical part, B. Fuks, O. Mattelaer, 
M. Selvaggi for correspondence on collider simulation, and F. Giordano, C. Leonidopulos for useful discussions on the CMS and ATLAS experimental features.   MN would like to thank B. Bajc, J.F. Kamenik and A. Urbano for useful discussions and E. Nemev\v sek for help on the manuscript. We would also like to thank G. Senjanovi\'c, V. Tello and Y. Zhang for helpful comments on the first version of the manuscript.   AM was supported in part by the 
Spanish Government and ERDF funds from the EU Commission  [Grants No. FPA2011-23778, No. CSD2007-00042 (Consolider Project CPAN)] and by Generalitat Valenciana under Grant No. PROMETEOII/2013/007.   MN was supported in part by the Slovenian Research Agency.  FN was partially supported by the Croatian Science Foundation (HrZZ) project "Physics of Standard Model and Beyond".

\end{acknowledgments}

\def\arxiv#1[#2]{\href{http://arxiv.org/abs/#1}{[#2]}}
\def\Arxiv#1[#2]{\href{http://arxiv.org/abs/#1}{#2}}

\end{document}